# On the 'insights into phases of liquid water from study of its unusual glass-forming properties'

Francesco Mallamace
*Dipartimento di Fisica, Università di Messina, Vill. S. Agata, C.P. 55, 98166 Messina, Italy*

Elpidio Tombari and Giuseppe Salvetti
*IPCF – CNRArea di Ricerca, Via G. Moruzzi 1, 56124 Pisa, Italy*

G. P. Johari*
*Department of Materials Science and Engineering, McMaster University, Hamilton, ON L8S 4L7, Canada*

(dated: 15 May 2008)

We investigate whether an interpretation of water's thermodynamics [Science, **319**, 582 (2008)] by using analogy with the binary metal alloys λ-type ordering transition or buckminsterfullerene's orientational-ordering transition has merit. On examining the heat capacity data used for the nanoconfined water, the construction of the heat capacity peak, and the number of water molecules in nanoconfinement, we find that (i) the peak had been obtained by joining the data for emulsified water with that of the nanoconfined water and (ii) only three water molecules can be fitted across the 1.1 nm diameter pores used in the study, two of which form a cylindrical shell that is hydrogen bonded to silica. The remaining connectedness of one water molecule would not produce a metal alloy-like λ-transition, or cooperative motions. Therefore, there is no basis for considering such an ordering in supercooled water.

To understand the experimental basis for the recently reported insights into water's thermodynamics and molecular kinetics [1], we examine how the plot of the total heat capacity, $C_p$, of the nanoconfined water's "internal component" in Fig. 2B of Angells' paper [1] had been obtained, particularly since no pore-size or other details were given in the paper. It is especially important to do so because the reconstructed 227 K peak in the $C_p$ against the temperature $T$ plot for supercooled water in Fig. 2B [1] was used as the only experimental basis for obtaining thermodynamic insight into supercooled water's behavior by seeking its analogy with the well-studied λ-type $C_p$ change at the disorder-order transition of binary metal alloys and of buckminsterfullerene [1]. We also determine what the pore size in these studies was, and then in what manner did the water molecules pack in nanoconfinement. An analysis of the calculated $C_p$ and other plots has been excluded, mainly because these lack immediate relevance and they used multiple approximations. Angell [1] cited two sources for the data in contructing Fig. 2B, Refs. 38 and 39. These are noted as [2] and [3] here.

First, we find that the reputed 227 K peak in Fig. 2B [1] had been produced by connecting the emulsified water's $C_p$ to the maximum $C_p$ value of the nanoconfined water measured until its value began to decrease by water's crystallization in silica gel. Murayama et al's [2] plot contains just one data point at the highest $T$ at which $C_p$ is slightly below the highest $C_p$ value measured. This data does not establish a $C_p$ peak at 227 K, because the measured $C_p$ of the nanoconfined water has at least an error of ± 1 %, and there is a considerable effect of rapid crystallization on the measured $C_p$ at the highest $T$. Therefore, the peak in Fig. 2B was not produced by the "total heat capacity of nanoconfined water" [1] alone. Rather, it had been reconstructed from the data in Fig. 2 of Murayama et al [2], which contained the nanoconfined water's $C_p$ only on the low temperature side of the reputed peak. There is also no evidence that $C_p$ of the nanoconfined water at higher $T$ would be the same as that of bulk water in 1.1 nm pore. Clearly, the experimental basis used for water's thermodynamic analysis [1] is unreliable.

We now consider the pore size for nanoconfined water whose $C_p$ data had been used [1]. It was given by Maruyama et al [2] as 3 nm. But this is not the case. The first suspicion comes from reading a detailed study by the same group [3] that stated: "Most of the water was found to crystallize within the pores above about 2 nm in diameter but to remain in the liquid state down to 80 K within the pores less than about 1.6 nm in diameter." This means that either water did not crystallize in the 3 nm pores of silica gel in Maruyama et al's study [2], or else the pore size of the silica gel was less than 1.6 nm.

It became evident in Oguni et al's Fig. 6a [3], that the actual pore size was less than 1.6 nm, as the plot in this figure had presented the same heat release data for the 1.1 nm pore water as had been reported earlier for the 3 nm pore water in Fig. 1a [2]. The revision for the pore size came after the nitrogen gas absorption/desorption isotherms in their Fig. 2b [3] showed that the average pore diameter in silica gel used in their earlier study was ~ 1.1 nm and not, as previously given, 3 nm. The silica-gel used by Maruyama et al [2] was later specified by Oguni et al [3] as silica Q-3 with 1.1 nm pore size. Therefore, the $C_p$ data plotted by Angell in his Fig. 2 B is for 1.1 nm confinement.

Since the reputed $C_p$ peak of the "internal component" of nanoconfined water in Fig. 2B [1] is the only experimental basis for a discussion of supercooled water's thermodynamics, we determine how the pore size effects the "internal component". It is widely known that water molecules at the silica interface form hydrogen bonds with the oxygen of the silica wall of the nanopore and they do not contribute to $C_p$ and other properties in the same manner as the remaining water molecules in the nanopore do. Maruyama et al [2] indeed stated: "The water confined within the pores are classified into two parts according to the regions in which it is located; internal water and interfacial water." Also: "The heat capacities of internal water within the 3 nm pores were derived by subtracting the contribution of interfacial water from the total of interfacial and internal water."[2]. Therefore, we need to determine first the number of water molecules that can be accommodated across the diameter of a nanopore and then the number of water molecules that can be regarded as "internal water". To do so, we estimate the volume of a water molecule as ~ 30 Å$^3$ (= 18/6.03 x10$^{23}$), and the effective diameter of its circumscribed sphere as 0.38 nm. Accordingly, only about eight water molecules (= 3.0/0.38) can fit across the diameter in a 3 nm pore, ignoring the excluded volume effect. Out of these, two would be interfacial molecules at the silica surface forming a cylindrical shell, which apparently does not crystallize [2], and six would be the "internal component" molecules, which crystallize. In Fig. 2B [1], this internal water's $C_p$ plot had been smoothly, albeit unjustifiably, joined with the $C_p$ plot for emulsified water.

As determined here earlier, the data in Fig. 2B [1] are for internal water confined in 1.1 nm pores. Therefore, we really need to consider the number of water molecules that can fit across the diameter of this size pore. Now, given the 0.38 nm diameter of a water molecule, only three (=1.1/0.38) water molecules can fit across the 1.1 nm pore diameter. Out of these, two would be interfacial water molecules bonded to the silica wall that form the cylindrical shell. This leaves only one water molecule to remain near the center of the 1.1 nm diameter cross section. This state of water molecules cannot be used for a thermodynamic discussion of bulk water. It is certainly not analogous to the ordering transition in metal alloys in which a λ-transition is observed and it makes little sense to speak of cooperative motions and glass transition in a three-layer thick nanoconfined water, with only one molecule layer not hydrogen bonded to silica.

Distortion of the $C_p$ – $T$ plots by varying the pore-size and/or other interfacial effects do not alter this conclusion, and the data used in constructing the $C_p$ peak in Fig. 2B has not been corrected for a possible distortion. Moreover, any distorted $C_p$ data plots would be unsuitable for combining with the emulsified or bulk water $C_p$ data. As an extension of their studies, Oguni et al [4] have now reported that $C_p$ of water confined to 1.2, 1.4 and 1.8 nm diameter linear pores in various types of MCM-41 (silica) varies with the pore size, particularly in the relevant range of 160 – 220 K. Their Fig. 2 shows that $C_p$ of confined water to 1.8 nm diameter pores of MCM-41 almost abruptly rises by more than a factor of two before its crystallization sets in. At first sight, it appears like a glass softening endotherm and, for that reason, Oguni et al [4] suggested that glass transition of bulk water occurs at 210 K, and not at 136 K or 165 K. They further suggested that supercooled water's relaxation time would perhaps still change from a non-Arrhenius to Arrhenius temperature dependence, but at $T$ not far from 210 K. This brings water's $T_g$ even closer to the questionable $C_p$-peak in Fig. 2B [1]. The new data appear to subvert the thermodynamic discussion provided by Angell [1].

This does not mean that the study of nanoconfined water is less meaningful than the study of other nanoconfined liquids; only that such studies need to be interpreted with care. This requires, a) avoiding a smooth graphical connection between the properties of nanoconfined water and emulsified or bulk water, or else assuming that they are the same, and, b) investigating the manner in which the water molecules can pack in nanopores. This is especially necessary because, (i) studies of the heat released on transfer of water *via* the vapor from the bulk state to confined state of 4 nm diameter pores in Vycor have clearly shown that the energy of a water molecule depends upon its position in a nanopore [5], (ii) $C_p$ is higher than for bulk water when the 4 nm pores are partly filled, and it decreases toward the bulk water value as they are gradually filled [6], and, (iii) the enthalpy and entropy of water in nanopores increase to the bulk water values as the amount of water in the nanopore is increased [7].

The $C_p$ plot of the 1.1 nm confined water in Fig. 2B [1] is the only experimental evidence given in the discussion intended to gain insights into the bulk water behavior. The above-given findings show that a reliable discussion of bulk water cannot be obtained from these data.

------------------------------------------------------------------------------------

*electronic mail: joharig@mcmaster.ca